%% file: manuscript.tex
\documentclass[preprint,12pt,authoryear]{elsarticle}

\usepackage{amssymb,amsmath,amsthm,multirow}
\usepackage[round]{natbib}
\newtheorem{definition}{Definition}
\newcommand{\sgn}{\text{sgn}}

\journal{Journal of Theoretical Biology}

\begin{document}

\begin{frontmatter}



\title{Estimating the probability of coexistence in cross-feeding communities}


\author{Bj\"{o}rn Vessman$^{*}$}
\author{Philip Gerlee}
\author{Torbj\"{o}rn Lundh}
\address{Mathematical Sciences, Chalmers University of technology and University of Gothenburg, 412 96 G\"{o}teborg}
\cortext[cor1]{Corresponding author e-mail: bvessman@mathbio.se}

\begin{abstract}
\input{Abstract}
\end{abstract}


\end{frontmatter}


\input{introduction.tex}
\input{3dsystem.tex}
\input{results.tex}

\input{discussion.tex}


\section{References}
\bibliographystyle{elsarticle-harv}
\bibliography{References}

\end{document}

%% file: Abstract.tex
The dynamics of many microbial ecosystems are driven by cross-feeding interactions, in which metabolites excreted by some species are metabolised further by others. The population dynamics of such ecosystems are governed by frequency-dependent selection, which allows for stable coexistence of two or more species. We have analysed a model of cross-feeding based on the replicator equation, with the aim of establishing criteria for coexistence in ecosystems containing three species, given the information of the three species' ability to coexist in their three separate pairs, i.e. the long term dynamics in the three two-species component systems. The triple-system is studied statistically and the probability of coexistence in the species triplet is computed for two models of species interactions. The interaction parameters are modelled either as stochastically independent or organised in a hierarchy where any derived metabolite carries less energy than previous nutrients in the metabolic chain. We differentiate between different modes of coexistence with respect to the pair-wise dynamics of the species, and find that the probability of coexistence is close to $\frac12$ for triplet systems with three pair-wise coexistent pairs and for so-called intransitive systems. Systems with two and one pair-wise coexistent pairs are more likely to exist for random interaction parameters, but are on the other hand much less likely to exhibit triplet coexistence. Hence we conclude that certain species triplets are, from a statistical point of view, rare, but if allowed to interact are likely to coexist. This knowledge might be helpful when constructing synthetic microbial communities for industrial purposes. 


%% file: introduction.tex
\section{Introduction}\label{secintro}

In recent years it has become increasingly clear that microbial species form complex communities, and rarely exist in isolation from each other \citep{Zelezniak2015}.  A common form of interaction occurs via the exchange of nutrients that are released by one species and absorbed and further metabolised by other species in the community.  This phenomenon is known as cross-feeding or synthropy and has been observed in a wide range of systems such as the human gut flora \citep{Belenguer2006}, the interactions of sulfate-reducers and methane oxidisers in the deep sea \citep{Hallam2004, Pernthaler2008}, the degradation of pesticides \citep{Katsuyama2009}, methanogenic environments \citep{stams}, and in soil nitrification \citep{Costa2006}. 

Note that cross-feeding can come in different degrees of complexity and interdependence. For example, when studying a system of {\it E. coli} strains feeding off an inflow of glucose, one strain is known to partially degrade the glucose to acetate, which would then be consumed by a second strain. Thus, the second strain will be affected by a negative frequency-dependent selection, as it needs the primary degrader. Furthermore, it has been put forward that the primary strain is dependent on the second one, as the secondary metabolite could be toxic at high concentrations, see for example \citep{pelz}.

Since the growth rate of a species within the community depends on the metabolites produced by other species, it indirectly depends on the frequency of other species. This implies that systems where cross-feeding is dominant are governed by frequency-dependent selection, which allows for both dominance and coexistence depending on the strength and sign of the interactions between the species \citep{Bomze1983}. 


Frequency-dependent selection (together with other mechanisms such as spatial structure \citep{Kerr2002}) is the most likely explanation for the stability of natural microbial ecosystems, such as the gut microbiota. Although we have mapped out a large number of microbial communities, we still lack a definite understanding of their dynamics and cannot explain why they are stable \citep{Harcombe2014}. This lack of knowledge becomes evident when it comes to assembling or constructing artificial communities that are stably maintained \citep{Jagmann2014}, and the need for understanding has become even more pressing now that the potential for engineering microbial communities for specific industrial purposes has been unravelled \citep{Grosskopf2014}. 

In order to build efficient microbial communities we could like to know if a given collection of species can form a stable ecosystem in which no species are outcompeted and driven to extinction. This question can be approached either from a top-down perspective using flux-balance analysis and co-occurrence data \citep{Zelezniak2015,vanHoek2016}, or from a detailed understanding of the population dynamics of cross-feeding systems. We take the latter approach and address the following straightforward and concrete question:
if we have qualitative information about the pair-wise dynamics of three species, what can be said about the likelihood of coexistence in the three-species community?




Thus, this paper strives to outline what configurations of species triplets that are likely to form coexistent populations. More specifically, we would like to know the coexistence properties of a triplet based on known pair-wise interactions between the constituent species. Ideally we would like to have quantitative information about the interactions of the different species, but for many practical purposes this is too much to ask. We therefore settle for qualitative information, and assume that for each pair of species we know if they coexist or if one species outcompetes the other. However, this poses a problem since it is known that systems that are identical on the pair-wise level (in the above qualitative sense), might behave differently when all three species are present  \citep{Bomze1983}.
This implies that triplet coexistence cannot be determined from the three pairs in isolation, but is a property of the interactions in the complete triplet. But all is not lost, since we may still be able to say something about the probability of coexistence. 
With this in mind we set out to study cross-feeding systems in which one, two or three pairs of species (out of the three) co-exist in isolation, and also intransitive systems where no dominant species exists (like rock-scissors-paper), and we do this from a statistical point of view to estimate the probability of coexistence in triplets of species.


\subsection{Mathematical modelling of cross-feeding}
The dynamics of large --- consisting of some billion cells or more --- and well-mixed populations of cross-feeding bacteria can be described by a system of coupled non-linear autonomous ordinary differential equations known as the replicator system of equations \citep{Lundh2013}. The assumption of well-mixedness allows us to disregard spatial effects in the system and in a large population, we may safely discard any stochastic individual interactions. The replicator system of equations has its origin in game theory \citep{Hofb2002} where it describes an evolutionary game of $n$ strategies and $d$ players \citep{Gokhale2010}, which corresponds 
to $n$ species and $d$ steps in the metabolic process in the cross-feeding framework. This correspondence is due to the fact that each metabolic step considered introduces an a coupled interaction between the species that take part in the metabolic chain.  

Replicator systems have been studied extensively \citep{Bomze1983, Hofb2002, Gokhale2010, Lundh2013} in relation to game theory and population dynamics, and we apply this theory to the present problem in order to find how pair-wise dynamics influence triplet coexistence. 


In the present setting, the fitness of a species is given by the amount of energy that the species can extract from the available nutrients excreted by another species. Here we interpret ``energy'' in a somewhat loose and abstract meaning. In this general setting the total energy is modelled as a sum over all possible metabolic interactions, represented as a series expansion of the fitness function. The model has been studied for two species by \citet{Lundh2013}, under the assumption that metabolites are only utilised by at most two species. In the same paper, the authors derived conditions for coexistence in a two-species population and the case of intransitive three-species populations. 

The dynamics of a cross-feeding ecosystem need not be modelled in the game-theoretical framework of the replicator system described by \citet{Lundh2013}. Other ODE-systems have described cross-feeding as a direct interaction between the involved species \citep{Bull2009, Estrela2010}, by explicitly modelling nutrient uptake and mortality \citep{Katsuyama2009}, and by using adaptive dynamics \citep{Doebeli2002}. Agent-based models have also been used by \citet{Gerlee2010} and \citet{Crombach2009}, whereas \citet{Pfeiffer2004} have studied the evolution of cross-feeding as a result of optimal ATP energy production in cells. In a recent study by \citet{Gedeon2015} cross-feeding in a chemostat environment was analysed. Under fairly general assumptions on the structure of the cross-feeding network they could show that there is a unique stable equilibrium that corresponds to the largest community of species that can be supported by the available resources, and that biomass production is maximised at this equilibrium point.

In this paper we take as a starting point the work of \citet{Lundh2013} and derive conditions for coexistence. Although we are able to derive analytical expressions for the conditions of coexistence, the large number of parameters in the model makes it difficult to draw any direct conclusions. Instead we approach the problem from a statistical point of view and randomly generate a large number of three species systems. The interaction parameters that model the energy uptake of a species are modelled in two ways: either as independent random variables or according to a hierarchical model where energy gains further down in the metabolic chain are lower than energy gains from primary metabolites. For a given parameter model, interaction parameters are drawn randomly from certain probability distributions to estimate the likelihood of permanence from the coexistence criteria.
Then, the relevant statistics of the sampled systems are computed and compared to the derived coexistence criteria.

\section{Preliminaries}\label{secPrelim}


The replicator system of equations for a population of species $i=1,\,2,\,\dots,\,n$ with individual frequencies $\mathbf{x}=(x_1,\,x_2,\,\dots,\,x_n)$ is defined as
\begin{equation}\label{repSystEq}
  \begin{cases}
    \dot{x}_i & = (\phi_i(\mathbf{x})-\bar{\phi}(\mathbf{x}))x_i, \\
    \bar{\phi}(\mathbf{x}) & = \sum\limits_{k=1}^nx_k\phi_k(\mathbf{x}),
  \end{cases}
\end{equation}
where $\dot{x}_i$ denotes the derivative with respect to time of a species frequency $x_i$, $\phi_i(\mathbf{x})$ is the species fitness function and $\bar{\phi}(\mathbf{x})$ is the average fitness in the population. Intuitively, a species that is fitter than the population average will increase in proportion to its current frequency and a species less fit than the average will decrease correspondingly. 

A fixed point of a system of ordinary differential equations $\dot{\mathbf{x}}=\mathbf{f(x)}$ is a point $\mathbf{x}^*$ in the domain of $\mathbf{f}$ such that $\mathbf{f}(\mathbf{x}^*)=0$. For the replicator system, the domain of definition for $\mathbf{f}$ is the simplex 
\begin{equation}\label{IntroNSimplex}
  S_{n-1} = \left\{\mathbf{x}\in\mathbb{R}^n|x_i\geq0, \ \sum\limits_{i=1}^n x_i = 1\right\},
\end{equation}
due to the requirement that the species frequencies are positive and defined as fractions of the whole population. The stability of the fixed points is determined \citep{Hofb2002} by the eigenvalues of the Jacobian 
\begin{equation}\label{eqJacobian}
J(\mathbf{x}^*) = 
  \left[  \frac{\partial f_i(\mathbf{x})}{\partial x_j}|_{\mathbf{x}=\mathbf{x}^*}
  \right]_{i,j}.
\end{equation}

Coexistence of species is related to the existence and location of the fixed points, but the system need not have stable fixed points in order to exhibit coexistence, and conversely the existence of stable fixed points does not imply coexistence. Rather, the property we are looking for is that of permanence, which is defined as:
\begin{definition}\label{defPermanence}
  A replicator system \eqref{repSystEq} is considered permanent if for all initial states $x_i(0)>0$, we have that $x_i(t)>0$ for all species $i=1,2,\dots,n$ and all $t>0$.
\end{definition}
The number of fixed points, their location and stability properties form the basis for the classification of solutions to the system \eqref{repSystEq}. Whether a fixed point is located on the boundary or in the interior of the domain of $\mathbf{f}$ is of special interest, since a permanent system is characterised by either a stable interior fixed point or a non-edge cyclic trajectory around a center fixed point. In the permanent case, we have $x_i>0$ for all species $i$ and thus that the interior fixed point $\mathbf{x}^*$ must satisfy $\phi_i(\mathbf{x}^*)-\bar{\phi}(\mathbf{x}^*)=0$ for all $i$, which means that the fitness of all species are equal
\begin{equation}\label{RepSystInnerFPCrit}
  \phi_1(\mathbf{x}^*) = \phi_2(\mathbf{x}^*) =\ ...\ = \phi_n(\mathbf{x}^*).
\end{equation}

The dynamics of the general three-species replicator system is studied and outlined in \citet{Bomze1983}.
The characterisation is based on the number of fixed points and their locations in the interior and boundary of the simplex. 
The fitness function used in that paper has the form
\begin{equation}\label{eqFitBomze}
  \phi_i= \sum\limits_{j=1}^n a_{ij}x_j,
\end{equation}
where the elements of a general normal-form payoff matrix are
\begin{equation}\label{PayoffZeroD}
  A = 
  \begin{cases}
     0, \ & i=j \\
     a_{ij}\in\mathbb{R}, \ & i\neq j
  \end{cases}.
\end{equation}
If a payoff matrix is not given in this zero-diagonal form, it may be transformed as such since the dynamics of the replicator system \eqref{repSystEq} does not change under column-wise addition of constants to the replicator system \citep{Bomze1983}.


\subsection{Replicator system for cross-feeding}\label{Sec3dRepOld}

For the cross-feeding model at hand, we recall here the derivation fitness function introduced by \cite{Lundh2013}, where it is assumed that the fitness of a species depends on its capacity to extract energy from a primary nutrient and from nutrients produced by other species (including itself). Three main assumptions were made in order to simplify the model: Firstly, it was  assumed that all uptake are equally efficient and linear with respect to the medium concentration, implicitly assuming that metabolites are scarce so that no saturation effects are present. The second assumption was that the amount of energy that can be extracted after the original metabolite being digested twice can be ignored. Thirdly, it was assumed that there is a separation in time scale between the dynamics of the metabolites and the bacterial population dynamics. 

\begin{figure}[!htb] 
  \centerline{ \includegraphics[width=0.55\textwidth]{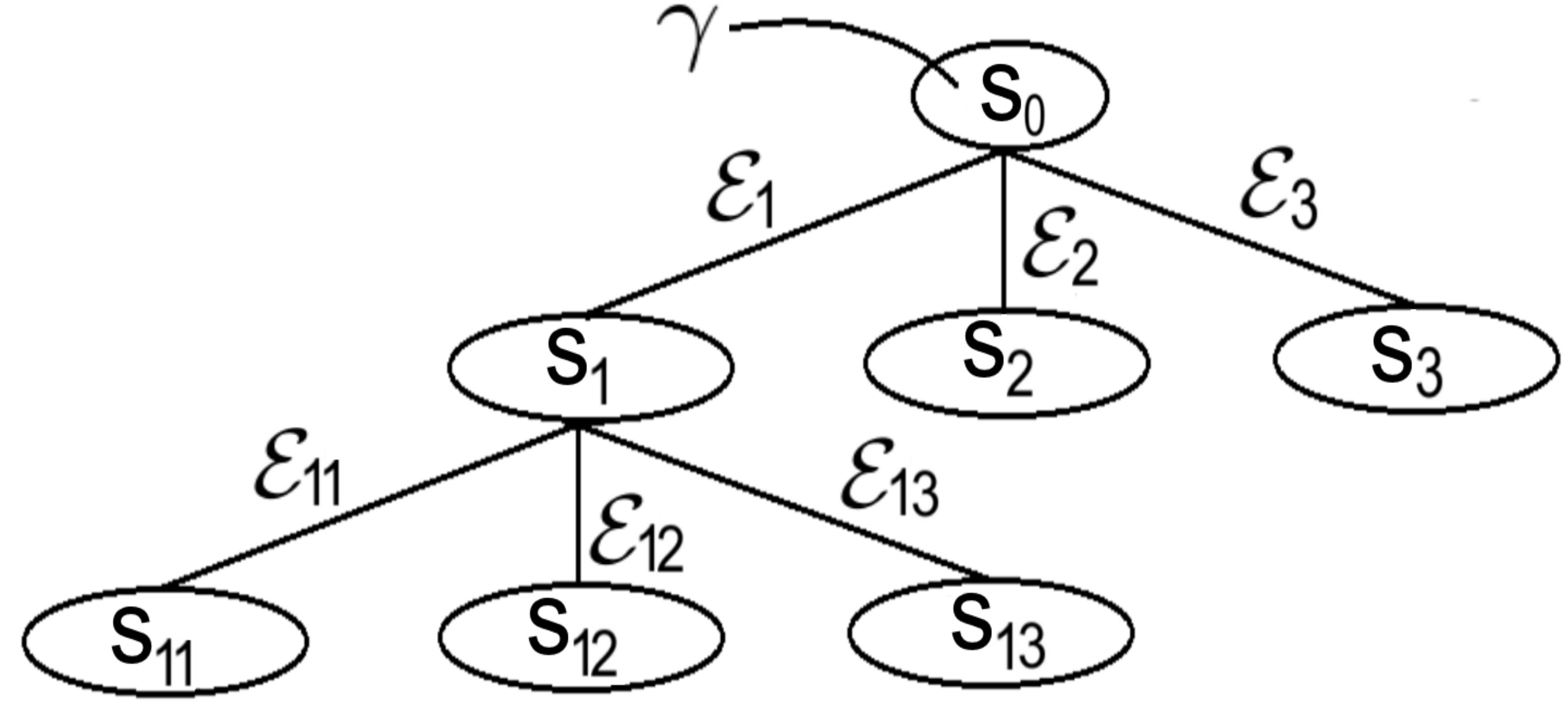} }
  \caption{First-order energy uptake  $\mathcal{E}_i$ for species $i$ from the primary nutrient $S_0$ that flows into the system at a rate $\gamma$, second-order metabolism $\mathcal{E}_{ij}$ for species $j$ from nutrient $S_i$. }\label{FigNutrientTree}
\end{figure}

For an illustration of the hierarchy of metabolites and energy extraction, see Figure~\ref{FigNutrientTree}. 

Species $i$ will gain an amount   $\mathcal{E}_i$ of energy when degrading the primary nutrient and $\mathcal{E}_{ji}$ when degrading metabolites excreted by species $j$. Since these parameters determine how the species interact with one another we term them \textit{interaction parameters}. To specify the interactions between three species we need $3+ 3 \times 3 = 12$ parameters. This number will later be revised, so that a payoff matrix on normal form only needs 9 interaction parameters. As by assumption two, any interactions of higher order than two are assumed negligible, so that the fitness of a species $i$ is approximated by the sum of two terms:
\begin{equation}\label{eq:fitness}
 \phi_i = \kappa s_0 x_i \mathcal{E}_i + \sum_{j} \kappa s_j x_i \mathcal{E}_{ji} 
\end{equation}
where $\kappa$ is a constant which describes the the effectiveness of uptake of the energy source. Thus the product $\kappa s_0 x_i$ in the model corresponds to the total uptake of the metabolite of bacteria from species $i$, and  $\gamma$ is the flow rate of metabolites into and out of the system.

This means that the concentration $s_0$ of the primary metabolite $S_0$ can be described by the following differential equation:
\begin{equation}\label{eq:S0}
\frac{ds_0}{dt}= - \sum_i \kappa s_0 x_i + \gamma (1-s_0)
\end{equation}
The concentration of a first order metabolite $s_i$ is then expressed as
\begin{equation}\label{eq:Si}
\frac{ds_i}{dt}= \kappa s_0 x_i - \sum_j \kappa s_i x_j - \gamma s_i.
\end{equation}

Since we are looking for a steady state, we let the right hand sides in  equations \eqref{eq:S0} and \eqref{eq:Si} be equal to equal to zero due to the third assumption above. This leads to a steady state of in primary resource $S_0$ at the level of $\frac{\gamma}{\kappa+\gamma}$, from which we define
\begin{equation}\label{uptakeToInflowEta}
  \eta = \frac{\kappa}{\kappa+\gamma}.
\end{equation}
Similarly, the steady-state level for $s_i$ is
\begin{equation}\label{eq:hatSi}
 \frac{\kappa\gamma x_i}{(\gamma+\kappa)^2} = \frac{\eta^2 \gamma x_i}{\kappa}.
\end{equation}
The steady state fitness of \eqref{eq:fitness} can thus be expressed as
\begin{equation}\label{fiteqLundh}
  \phi_i(\mathbf{x}) = \eta\gamma\mathcal{E}_i +\eta^2\gamma\sum\limits_j\mathcal{E}_{ji}x_j.
\end{equation}
Note that this equation is a simplification of \cite[Eq. (9)]{Lundh2013} with the higher order terms ignored. 

In a three-species replicator system, there is the possibility of stable interior fixed points as well as stable and unstable fixed points anywhere on the boundary of the system. \citet{Bomze1983,Bomze1995} characterises no less than 49 different types of phase portraits for a three-species replicator system, with the main division being the number of fixed points in the interior of the simplex  \citep{Bomze1983} 
\begin{equation}\label{3dSimplex}
  S_2 = \left\{x\in\mathbb{R}^3| x_i\geq0,\ \sum\limits_{i=1}^3x_i=1\right\}.
\end{equation}




\subsection{Models for interactions parameters}\label{ssec2dPermProb}
The interactions parameters $\mathcal{E}_i$, $\mathcal{E}_{ij}$ are considered as random variables and modelled in two distinct ways. In the first scheme, $\mathcal{E}_i$ and $\mathcal{E}_{ij}$ are assumed to be independent in the stochastic sense, and drawn from a uniform distribution on the unit interval, i.e. $\mathcal{E}_i \sim\text{Uni}(0,1)$ and $\mathcal{E}_{ij} \sim\text{Uni}(0,1)$ for all species $i$ and $j$. The uniform distribution is chosen due to its simplicity and can be seen to represent no prior knowledge on the interaction parameters. Please note that the positivity of the interaction terms (as they represent an energy gain) restricts us from using a normal distribution, which otherwise would have been a natural choice \citep{Gokhale2010}. 

In the second scheme we constrain the amount of energy extracted at higher levels of cross-feeding by assuming that it is necessarily smaller than the amount extracted from the primary resource. This implies that for a fixed species $i$, we have 
\begin{equation}\label{treeProperty}
  \mathcal{E}_i > \mathcal{E}_{ij} > 0
\end{equation} 
for all species $j$. We implement this by letting $\mathcal{E}_i \sim\text{Uni}(0,1)$ and defining the second order terms as
\begin{equation}\label{treeHierarch2}
  \mathcal{E}_{ij}=r_{ij}\mathcal{E}_i 
\end{equation}
where the scaling factor $r_{ij}$ is drawn from a $\text{Uni}(0,1)$-distribution. 

In order to test the generality of our results we also consider the two schemes with exponentially distributed parameters with intensity $2$ (having mean 1/2, the same as the $\text{Uni}(0,1)$-parameters).

%% file: 3dsystem.tex
\section{Analytical results}\label{Sec3dRepNew}
We are now ready to discuss how the results of \citet{Bomze1983} and \citet{Lundh2013} can be used to determine the probability of permanence for triplet systems. First, we will outline which of the previous results that are of interest and how they relate to triplet coexistence, as well as define a necessary condition for permanence in Section~\ref{Ssec3dPermGeneral}. Then, in Sections~\ref{Ssec3dCyclic}--\ref{Ssec3dIntrans} we will describe the types of pair-wise interactions in the ecosystem that are of interest and how to define each of them. 


In order to compare the replicator systems proposed by \citet{Lundh2013} to that of \citet{Bomze1983}, we may use a technique described by \citet{Gerstung2011} and \citet{Stadler1990}, namely that we define an alternative payoff matrix $E$ with elements
\begin{equation}\label{eq3dEiEijMatrix}
  E_{ji} = \gamma\eta\mathcal{E}_i +\gamma\eta^2\mathcal{E}_{ji}
\end{equation}
so that the fitness function \eqref{fiteqLundh} may be written as
\begin{equation}\label{eq3dESystem}
  \widetilde{\phi}_i(\mathbf{x}) = \sum\limits_{j=1}^3E_{ij}x_j
\end{equation}
which is equivalent to the linear fitness function \eqref{eqFitBomze}. 
The proof of equivalence is straightforward and relies on the fact that $\sum_{j=1}^3x_j = 1$ \citep{Gerstung2011}. 

By the property mentioned in Section~\ref{secPrelim} that a payoff matrix may be transformed by column-wise addition and subtraction, we will use an alternate form of \eqref{eq3dEiEijMatrix} to be able to directly use the results of \citet{Bomze1983}, namely:
\begin{equation}\label{3dLGPayoff2}
  E = \gamma\eta
  \begin{bmatrix}
    0 & 0 & 0 \\
    \lambda_{12}-\lambda_{11} & \lambda_{22}-\lambda_{21} & \lambda_{32}-\lambda_{31} \\
    \lambda_{13}-\lambda_{11} & \lambda_{23}-\lambda_{21} & \lambda_{33}-\lambda_{31}
  \end{bmatrix},
\end{equation}
where we have the definition of total energy uptake
\begin{equation}\label{3dIntransLambdaRepeat}
  \lambda_{ji} = \mathcal{E}_i +\eta\mathcal{E}_{ji}.
\end{equation}

\subsection{Stable interior fixed points}\label{Ssec3dPermGeneral}

\begin{figure}[!htb]
\begin{center}
\includegraphics[width=6cm]{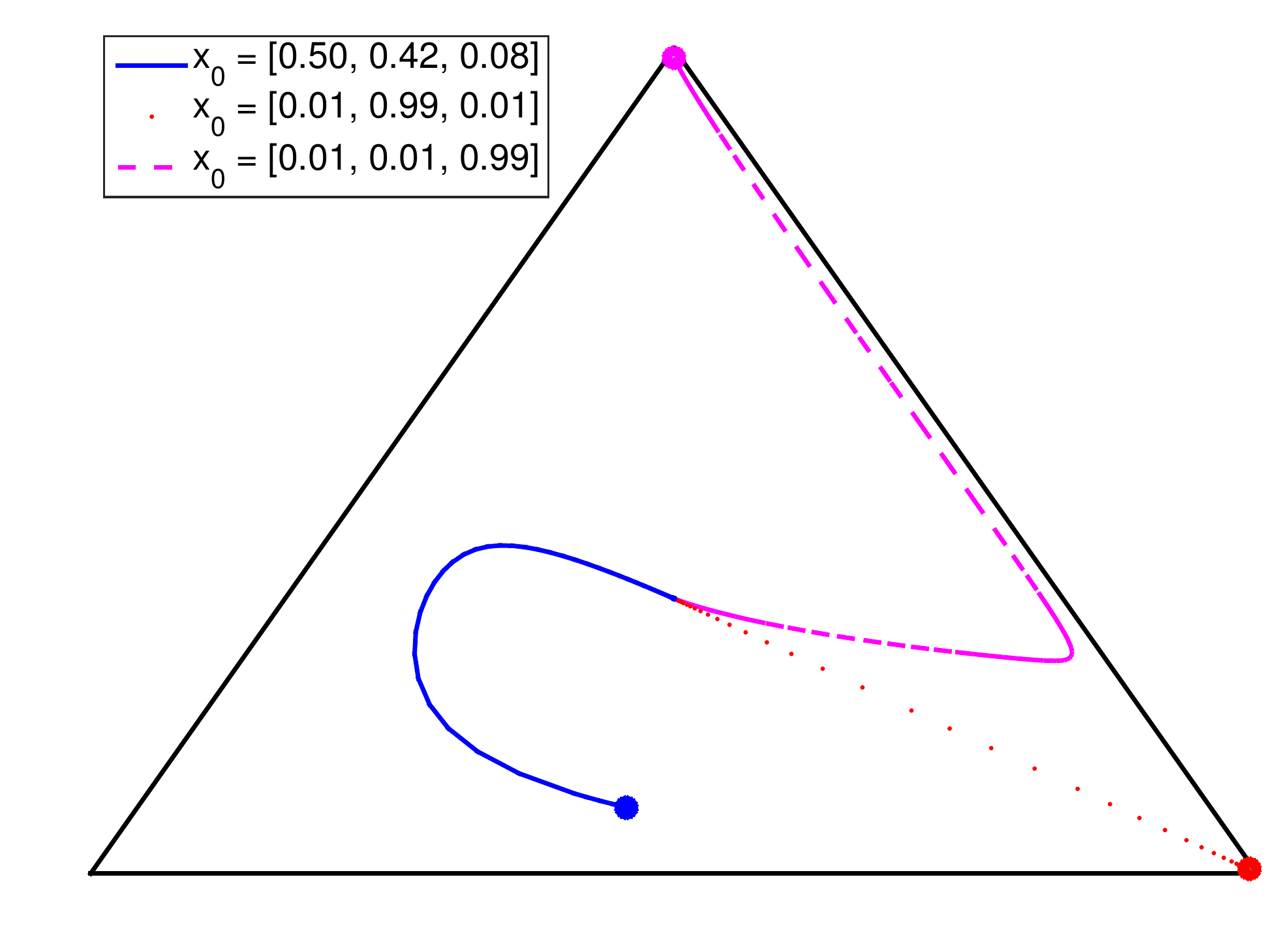}
               \includegraphics[width=6cm]{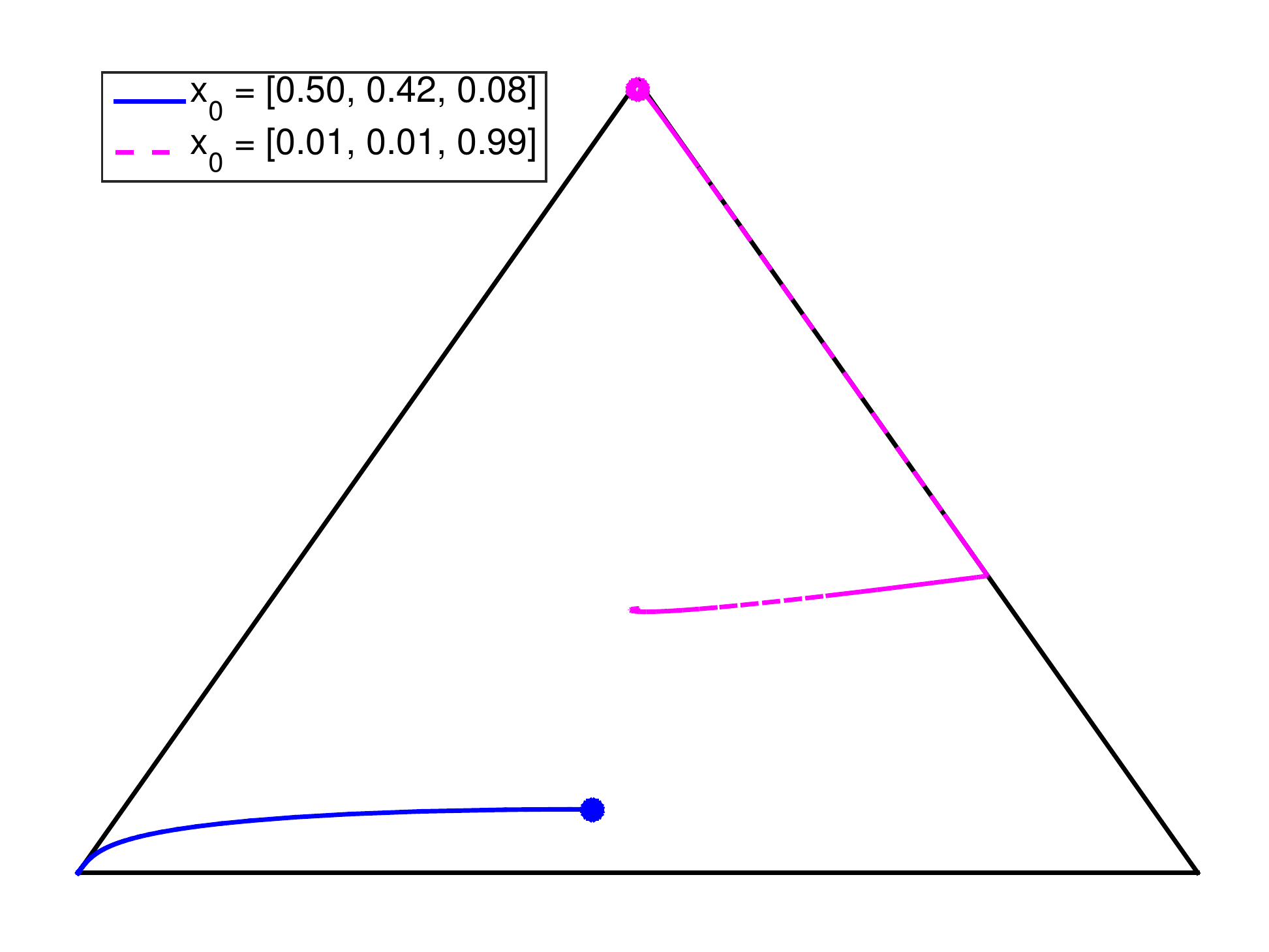}
\caption{\label{FigPermVSCondPerm} Example of a) permanent and b) conditionally permanent system. The dots mark the initial configuration of the system, as shown in the legend. In the permanent case all initial conditions converge to the same stable steady state, whereas in the conditionally permanent system the two initial conditions converge to different steady states with different stability properties. }
\end{center}
\end{figure}

We are now to derive general conditions for existence and stability of fixed points (FPs) in the interior of the state space $S_2$, defined as \eqref{3dSimplex}. 
A stable interior fixed point is not a sufficient condition for permanence, as there are replicator systems with a stable fixed point in the interior of the simplex that cannot be reached from all initial states. These systems will be called conditionally permanent, and an example of trajectories in permanent vs.\ conditionally permanent systems is shown in Figure~\ref{FigPermVSCondPerm}. In biological terms, conditional permanence means that a species triplet may be stable at certain initial frequencies but not at others. Dynamical systems sometimes exhibit stable trajectories in so-called limit cycles, where solutions tend towards a cyclic trajectory where all frequencies are non-zero. For a replicator system of less than four species, no such limit cycles are possible, as proven by \citet{Hofbauer1981}.

The method we use is due to \citet{Bomze1983} and also used by \citet{Stadler1990} and is based on finding two coordinates $p,\ q$ that define a unique fixed point 
\begin{equation}\label{3dGeneralInnerFP}
  \mathbf{x}^*=\frac{1}{1+p+q}(1,\ p,\ q).
\end{equation}
that lies in the interior of $S_2$ when $p,\ q$ are positive. The coordinates are given by
\begin{align}
  p = \frac{\Delta_1}{\Delta_3} \label{Bomze_pq1}\\
  q = \frac{\Delta_2}{\Delta_3} \label{Bomze_pq2}
\end{align}
where $\Delta_k$ are the $2\times2$ co-factors of the payoff matrix \eqref{3dLGPayoff2}, defined as
\begin{align}
  \Delta_1 &= (\lambda_{32}-\lambda_{31})(\lambda_{13}-\lambda_{11})-(\lambda_{12}-\lambda_{11})(\lambda_{33}-\lambda_{31}) \label{BomzePQSubd1}\\
  \Delta_2 &= (\lambda_{12}-\lambda_{11})(\lambda_{23}-\lambda_{21})-(\lambda_{22}-\lambda_{21})(\lambda_{13}-\lambda_{11}) \label{BomzePQSubd2}\\
  \Delta_3 &= (\lambda_{22}-\lambda_{21})(\lambda_{33}-\lambda_{31})-(\lambda_{32}-\lambda_{31})(\lambda_{23}-\lambda_{21}) \label{BomzePQSubd3}
\end{align}
where $\lambda_{ji}=\mathcal{E}_{i}+\eta\mathcal{E}_{ji}$.

The criterion for existence of the fixed point \eqref{3dGeneralInnerFP} is that $p,\ q$ are real and positive, which occurs when all of the $\Delta_i$ have the same sign,
\begin{equation}\label{3dSignCrit}
  \sgn(\Delta_1)=\sgn(\Delta_2)=\text{sgn}(\Delta_3)
\end{equation}
where $\sgn\vert\mathbb{R}\rightarrow\{-1,\ 0,\ 1\}$ denotes the sign function. The stability properties of the fixed point is determined by the Jacobian's \eqref{eqJacobian} eigenvalues 
\begin{equation}\label{3dEigsBomzeCorr6} 
  \mu_{1,2}= \frac{1}{2}\left( \alpha p+\beta q \pm\sqrt{(\alpha p+\beta q)^2-4pq\Delta_3}\right)
\end{equation}
where we have defined, for notational convenience, 
\begin{align}
  \alpha &= \lambda_{22}-\lambda_{21}\label{alphaFactorEig}\\
  \beta  &= \lambda_{33}-\lambda_{31}\label{betaFactorEig}
\end{align}
The system is stable when both eigenvalues are real and negative or when the real part of the complex eigenvalues are negative, i.e., when 
\begin{equation}\label{3dEigsBomzeCorr7Crit} 
  \alpha p +\beta q <0.
\end{equation}
In conclusion the conditions for existence and stability of fixed points in the interior of the simplex are (i) that the co-factors have the same sign \eqref{3dSignCrit} and (ii) that the real part of the eigenvalues are negative \eqref{3dEigsBomzeCorr7Crit} (see Table~\ref{Tab3dGenPerm}). 


\begin{table}[!htb]
  \centering
  \begin{tabular}{|c|c|c|}\hline
    Inner fixed point        & \multirow{2}{*}{Existence \eqref{3dSignCrit}}          & Stability \\
    \eqref{3dGeneralInnerFP} &  & \eqref{3dEigsBomzeCorr7Crit} \\\hline
    $\mathbf{x}^*$ & $\sgn(\Delta_1)=\sgn(\Delta_2)=\sgn(\Delta_3)$ & $\alpha p +\beta q <0$ \\\hline
  \end{tabular}
  \caption{ Criteria for permanent system with a stable fixed point in the interior of the simplex. \citep{Bomze1983} }\label{Tab3dGenPerm}
\end{table}

\subsection{Centre fixed points and cyclic trajectories}\label{Ssec3dCyclic}

\begin{figure}[h] 
  \centerline{ 
               \includegraphics[width=0.65\textwidth]{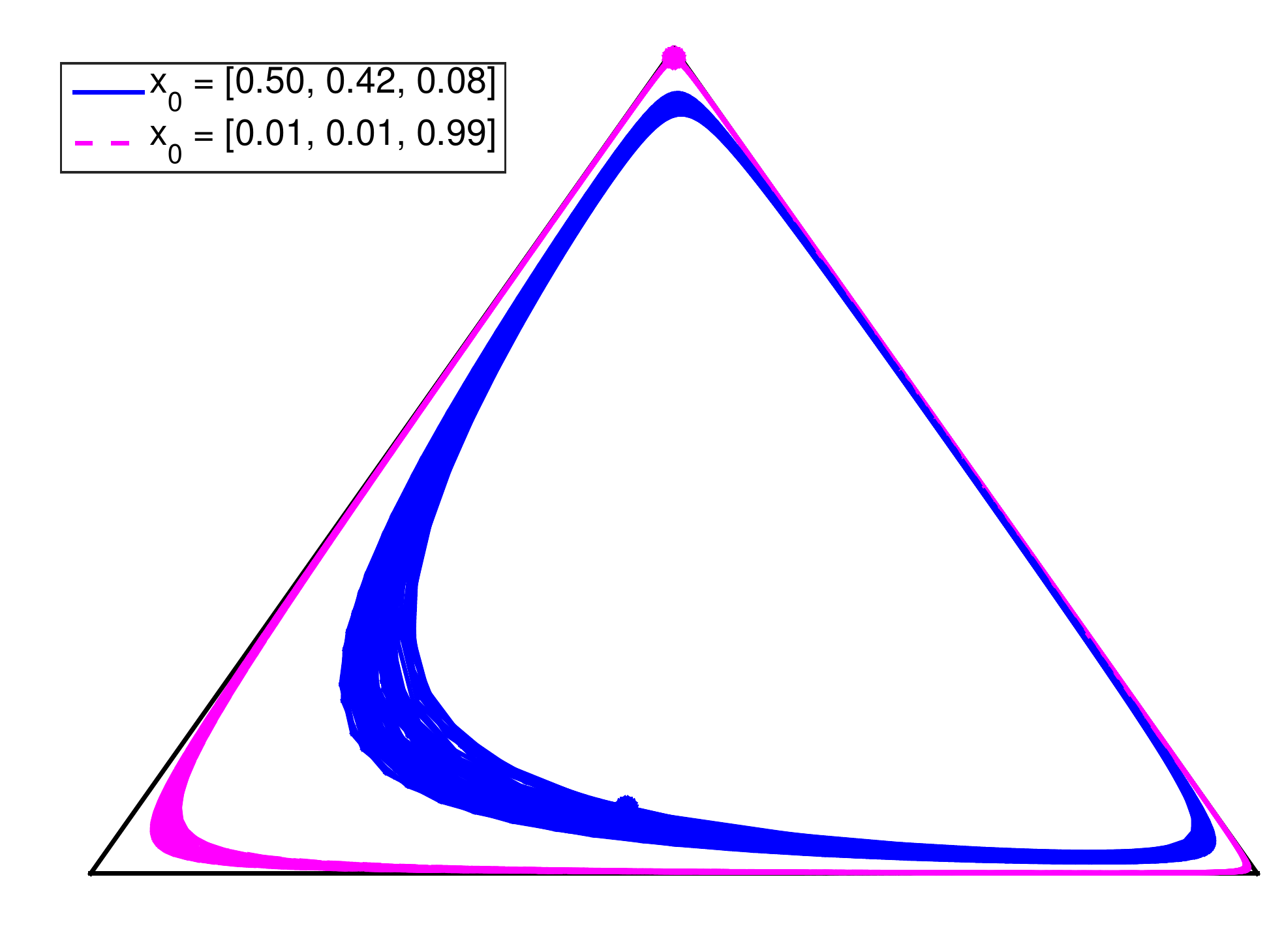} 
              }
   \caption{Cyclic rock-paper-scissors replicator system (solid blue curve, inner loop) compared to system with stable cyclic fixed point (dashed magenta, outer loop) with negative real part of the eigenvalues \eqref{3dEigsBomzeCorr6}.}\label{figRPS3d}
\end{figure}

We noted before that permanence might be the result of cyclic trajectories in the interior of the state space. It is therefore valuable to know for which systems this occurs. 
If the eigenvalues to the payoff matrix of the replicator system are strictly imaginary, i.e. with a zero real part, in a neighbourhood of an inner fixed point, then all trajectories in the neighbourhood will be cyclic and the fixed point is a center. An example of a center system is shown in Figure~\ref{figRPS3d}.


The center fixed point is the limiting case between complex eigenvalues with negative and positive real parts. 
However, from a probabilistic point of view, we may note that strictly imaginary eigenvalues require an exactly zero real part of \eqref{3dEigsBomzeCorr6}. Since all terms depend on the random variables $\mathcal{E}_i$ and $\mathcal{E}_{ij}$, we have that equality constitutes a zero-measure set and hence we expect a zero probability of finding such systems from randomly generated parameters. 

\subsection{Pair-wise coexistence}\label{Ssec3dFriends}
We have now set out the general criteria for triplet coexistence and will now investigate how pair-wise behaviour affects coexistence. In a permanent three-species system, it may be the case that the involved species are pair-wise coexistent when isolated from the third species. In terms of dynamical systems, the criterion for pair-wise coexistence means that there exists one or more stable fixed points on the non-corner edges of the simplex. If the fixed point on a given edge is stable, any trajectories along that edge will converge to the fixed point as $t\rightarrow\infty$. 

In the interior of the simplex near a stable boundary fixed point there are however two possible behaviours: a semi-stable fixed point will repel the trajectories in the part of its neighborhood that lies in the interior of the simplex, whereas a stable fixed point will attract any trajectory in its neighborhood.   

We now analyse the situation where we have pair-wise coexistence of one or more of the three species pairs, with a method described by \citet{Stadler1990}. 
 These boundary fixed points must be semi-stable, and hence attract trajectories on the boundary and repel trajectories in the interior of the simplex. To analyse this situation we use the payoff matrix \eqref{eq3dEiEijMatrix} on its normal form 
\begin{equation}\label{3dLGPayoffNormal}
  E = \gamma\eta
  \begin{bmatrix}
    0                         & \lambda_{21}-\lambda_{22} & \lambda_{31}-\lambda_{33} \\
    \lambda_{12}-\lambda_{11} & 0                         & \lambda_{32}-\lambda_{33} \\
    \lambda_{13}-\lambda_{11} & \lambda_{23}-\lambda_{22} & 0
  \end{bmatrix},
\end{equation}
that corresponds the homogenous form of the replicator system. We find that the fixed points on the non-corner edge of the simplex are
\begin{align}
  \mathbf{x}^*_{12} = \frac{1}{\lambda_{12}-\lambda_{11}+\lambda_{21}-\lambda_{22}}(\lambda_{21}-\lambda_{22},\ \lambda_{12}-\lambda_{11},\ 0) \label{3dFriendFPsA}\\
  \mathbf{x}^*_{23} = \frac{1}{\lambda_{23}-\lambda_{22}+\lambda_{32}-\lambda_{33}}(0,\ \lambda_{32}-\lambda_{33},\ \lambda_{23}-\lambda_{22}) \label{3dFriendFPsB}\\
  \mathbf{x}^*_{31} = \frac{1}{\lambda_{13}-\lambda_{11}+\lambda_{31}-\lambda_{33}}(\lambda_{31}-\lambda_{33},\ 0,\ \lambda_{13}-\lambda_{11}) \label{3dFriendFPsC}
\end{align}
under the conditions
\begin{align}
  (\lambda_{12}-\lambda_{11})(\lambda_{21}-\lambda_{22})>0 \label{3dFriendFPsCondA} \\
  (\lambda_{23}-\lambda_{22})(\lambda_{32}-\lambda_{33})>0 \label{3dFriendFPsCondB} \\
  (\lambda_{13}-\lambda_{11})(\lambda_{31}-\lambda_{33})>0 \label{3dFriendFPsCondC}
\end{align}
which ensure that each pair of payoff elements, for example $\lambda_{21}-\lambda_{22}$ and $\lambda_{12}-\lambda_{11}$ of \eqref{3dFriendFPsA}, have the same sign so that the coordinates of $\mathbf{x}^*_{12}$ are properly defined on the intervals $[0,1)$ when normalised. 

In order to have permanence, we require that the edge fixed points are semi-stable fixed points which attract trajectories on the edge and repel trajectories in the interior of the simplex, so-called saddle points. The previous conditions hold for both stable and unstable fixed points, and to single out the fixed points which are stable along the edges, we modify the conditions \eqref{3dFriendFPsCondA}-\eqref{3dFriendFPsCondC} to require positiveness for each element in the pairs of payoff elements
\begin{align}
  (\lambda_{12}-\lambda_{11})>0,\ (\lambda_{21}-\lambda_{22})>0 \label{3dFriendFPsCondA2}\\
  (\lambda_{23}-\lambda_{22})>0,\ (\lambda_{32}-\lambda_{33})>0 \label{3dFriendFPsCondB2}\\
  (\lambda_{13}-\lambda_{11})>0,\ (\lambda_{31}-\lambda_{33})>0 \label{3dFriendFPsCondC2}
\end{align}
so that edge-bound trajectories on both sides of the fixed point will tend to the fixed point \citep{Stadler1990}. We will now discuss the cases where three, two and one fixed points exists on the boundary of the system. Biologically this corresponds to situations where three, two and one of the three pairs exhibit stable coexistence in isolation. 



In the case of pair-wise coexistence of all three species we have existence of unique fixed points $\mathbf{x}^*_{ij}$ as of \eqref{3dFriendFPsA}-\eqref{3dFriendFPsC} on the edges of the simplex when
the conditions \eqref{3dFriendFPsCondA}-\eqref{3dFriendFPsCondC} are fulfilled. Furthermore, the fixed points are semi-stable when there exists a stable interior fixed point and the conditions \eqref{3dFriendFPsCondA2}-\eqref{3dFriendFPsCondC2} hold (Table \ref{Tab3dGenPerm}). We note that the criteria for semi-stability of the boundary fixed points imply existence of the interior fixed point. The conditions are collected in Table~\ref{tabFriendsConds}, where all conditions are necessary for a system with three pair-wise fixed points and one stable triplet fixed point in the interior of the state space.
\begin{table}[!htb]
  \centering
  \begin{tabular}{|c|c|c|c|}\hline
    Fixed point & Existence & \multicolumn{2}{c|}{ Stability } \\\hline
     $\mathbf{x}_{12}^*$ \eqref{3dFriendFPsA} & $(\lambda_{12}-\lambda_{11})(\lambda_{21}-\lambda_{22})>0$ \eqref{3dFriendFPsCondA} & $(\lambda_{12}-\lambda_{11})>0$ & $(\lambda_{21}-\lambda_{22})>0$ \\\hline
     $\mathbf{x}_{23}^*$ \eqref{3dFriendFPsB} & $(\lambda_{23}-\lambda_{22})(\lambda_{32}-\lambda_{33})>0$ \eqref{3dFriendFPsCondB} & $(\lambda_{23}-\lambda_{22})>0$ & $(\lambda_{32}-\lambda_{33})>0$ \\\hline
     $\mathbf{x}_{31}^*$ \eqref{3dFriendFPsC} & $(\lambda_{31}-\lambda_{33})(\lambda_{13}-\lambda_{11})>0$ \eqref{3dFriendFPsCondC} & $(\lambda_{31}-\lambda_{33})>0$ & $(\lambda_{13}-\lambda_{11})>0$ \\\hline
     $\mathbf{x}^*$ \eqref{3dGeneralInnerFP} & $\sgn(\Delta_i)=\sgn(\Delta_{j})\ \forall i,\ j$ \eqref{3dSignCrit} & \multicolumn{2}{c|}{ $\alpha p +\beta q <0$ \eqref{3dEigsBomzeCorr7Crit} } \\\hline
  \end{tabular}
  \caption{ Criteria for permanent system with semi-stable fixed points along boundary. }\label{tabFriendsConds}
\end{table}

For a system with two pairs that are coexistent in isolation from the third species, we require that any two of the conditions of edge fixed points in Table~\ref{tabFriendsConds} hold. This corresponds to the case where one of the edge fixed points have migrated to a corner of the simplex when payoff entries have different signs. Recall that a stable fixed point in the interior of the simplex is not in itself a sufficient condition for permanence, as there are initial states that do not converge to the fixed point. 

Similarly, for a system with one coexistent pair, only one of the fixed points \eqref{3dFriendFPsA}-\eqref{3dFriendFPsC} exists on the non-corner edge of the simplex.


\subsection{Intransitivity and permanence}\label{Ssec3dIntrans}
A special type of pair-wise relation, which is known to promote coexistence \citep{Kerr2002}, is that of an intransitive species triplet. In this case the species dominate each other in a circular fashion, just as in the game rock-scissors-paper. This can be expressed as
\begin{equation}\label{3dIntransCond}
  \big(\lambda_{i+1,i}-\lambda_{i+1,i+1}\big)\big(\lambda_{i,i}-\lambda_{i,i+1}\big) > 0.
\end{equation}
where $\lambda_{ij}$ is the $i$-th row, $j$-th column element of the payoff matrix \eqref{eq3dEiEijMatrix}, and we consider all indices modulo $3$. This condition follows from considering the ordered pairs $(i,i+1)$ such that the species $i+1$ outcompetes species $i$ when no other species are present. In terms of the phase portrait of the replicator system, this means that the fixed point $(x^*_{i},x^*_{i+1})=(1,0)$ is unstable and that the fixed point $(x^*_{i},\ x^*_{i+1})=(0,1)$ is stable. 

The criterion for permanence for an intransitive three-species replicator system is
\begin{equation}\label{3dCoexCrit}
  \Gamma_{12}\Gamma_{23}\Gamma_{31}<1,
\end{equation}
as stated in Theorem~2 of \citet{Lundh2013}. The permanence factors $\Gamma_{ij}$ are defined as
\begin{equation}\label{3dGammaIJ}
  \Gamma_{ij} = \frac{\lambda_{ji}-\lambda_{jj}}{\lambda_{ii}-\lambda_{ij}},
\end{equation} 
and are positive as long as the system is permanent per condition \eqref{3dIntransCond}. 
In conclusion, the three-species replicator system \eqref{repSystEq} is pair-wise intransitive and permanent if condition \eqref{3dCoexCrit} holds together with \eqref{3dIntransCond}. The conditions are collected in Table~\ref{tabIntransConds}.

\begin{table}[!htb]
  \centering  
  \begin{tabular}{|l|c|}\hline
    Species pair & $(i, i+1)\mod 3$ \\\hline
      Intransitive if \eqref{3dIntransCond}
      & $(\lambda_{i+1,i}- \lambda_{i+1, i+1})(\lambda_{i, i}-\lambda_{i, i+1}) >0$ \\
    Permanent if \eqref{3dCoexCrit} & $\Gamma_{12}\Gamma_{23}\Gamma_{31}<1$ \\\hline
  \end{tabular}
  \caption{ Criteria for existence of intransitive triplet. }\label{tabIntransConds}
\end{table}


%% file: results.tex
\section{Numerical evaluation of permanence criteria}\label{SecNumRes}
For a general three-species system, the derived criteria for existence and stability of fixed points in the 2-simplex are hard to interpret due to the high dimensionality of the parameters (9 interaction parameters in total). We therefore investigate them from a probabilistic point of view rather than study them analytically. The statistical properties of three-species systems are evaluated numerically for random interactions parameters drawn according to the Uni(0,1)-distribution in the independent and hierarchical parameter model (see section \ref{ssec2dPermProb} for details). We also extend the analysis to the case where the interaction parameters are drawn from an exponential distribution (with mean $1/2$). The results were averaged across $N=10^6$ different realisations (independent draws of the interaction parameters) and we used parameter values $\gamma = 0.03$ and $\kappa=0.25$ \citep{Lundh2013}. The empirical probabilities of finding a system with a certain property are estimated as the fraction of systems that satisfy the conditions (equalities and/or inequalities) that correspond to the property.

\subsection{Stable interior fixed point}\label{Res3dGeneral}

Existence of the interior fixed point $\mathbf{x}^*=\frac{1}{1+p+q}(1,\ p,\ q)$ is determined by the signs of the sub-determinants \eqref{BomzePQSubd1}-\eqref{BomzePQSubd3}, which gives the criterion
\begin{equation}\label{Res3dGenPermExist}
  \sgn(\Delta_1)=\sgn(\Delta_2)=\sgn(\Delta_3).
\end{equation}
The necessary criterion for stability of the fixed point is negative real parts of the eigenvalues \eqref{3dEigsBomzeCorr6} to the interior fixed point, which is ensured by 
\begin{equation}\label{3dEigsBomzeCorr7CritRep} 
  \alpha p +\beta q <0.
\end{equation}
Table~\ref{tab3dGenPerm} collects the results from the independent and hierarchical model of the interaction parameters for uniformly distributed random variables. We note that for both models the probability of existence of a fixed point is on the order of 1\% and the probability that this fixed point is also stable is roughly one order of magnitude smaller. 

\begin{table}[!htb]
  \centering
  \begin{tabular}{cc|c|c|}\hline
    \multicolumn{2}{|c|}{\multirow{2}{*}{Probability}} 
                 & Pr(FP exists) & Pr(Stable FP exists) \\ 
    \multicolumn{2}{|c|}{ } & Eq.\ \eqref{Res3dGenPermExist} & Eq.\ \eqref{Res3dGenPermExist},\eqref{3dEigsBomzeCorr7CritRep}  \\ \hline
    \multicolumn{1}{ |c  }{\multirow{2}{*}{$\mathcal{E}_i$ model} } &
    \multicolumn{1}{ |l| }{Uni$(0,1)$, Indep.} & 7.1$\times 10^{-2}$ & 18$\times 10^{-3}$\\ \cline{2-4}
    \multicolumn{1}{ |c  }{}                        &
    \multicolumn{1}{ |l| }{Uni$(0,1)$, Hier.} & 1.2$\times 10^{-2}$ & 3.2$\times 10^{-3}$\\ \cline{1-4}
  \end{tabular}
  \caption{ Empirical probability of existence of a stable interior fixed point (FP). }\label{tab3dGenPerm}
\end{table}

\subsection{Pair-wise coexistence}\label{Res3dPWCoex}

For a pair-wise coexistent triplet, we require one, two or three semi-stable fixed points on the non-corner boundary of the simplex (in addition to a stable interior fixed point). The criteria for semi-stability that are found in Table~\ref{tabFriendsConds} are sufficient also for existence of the boundary fixed points. 

\begin{table}[!htb]
  \centering
  \begin{tabular}{cc|c|c|c|}\hline
    \multicolumn{2}{|c|}{ Pr($n$ semi-stable FPs on boundary) } & $n=3$ & $n=2$ & $n=1$ \\ \hline
    \multicolumn{1}{ |c  }{\multirow{2}{*}{$\mathcal{E}_i$ model} } &
    \multicolumn{1}{ |l| }{Uni$(0,1)$, Indep.} & 12$\times10^{-3}$ & 6.0$\times10^{-2}$ & 2.2$\times10^{-1}$ \\\cline{2-5}
    \multicolumn{1}{ |c  }{}                        &
    \multicolumn{1}{ |l| }{Uni$(0,1)$, Hier.} & 2.2$\times10^{-3}$ & 1.2$\times10^{-2}$ & 1.4$\times10^{-1}$ \\ \cline{1-5}
  \end{tabular}
  \caption{ Empirical probability of coexistence in systems with three, two or one stable fixed points \eqref{3dFriendFPsA}-\eqref{3dFriendFPsC} on the boundary, by criteria in Table~\ref{tabFriendsConds}. }\label{tab3d123Friends}
\end{table}

We note that the probability of coexistence is decreasing with the number of required fixed points on the boundary. This is not surprising since more inequalities need to hold for two and three pairs.

\subsection{Intransitivity and permanence}\label{SecResIntrans}

Existence of intransitive systems is determined by the conditions \eqref{3dIntransCond}, that need to hold for all pairs $(\mathbf{x}_i,\ \mathbf{x}_{i+1})$ where we as usual consider the indices modulo 3. The results are collected in Table~\ref{tab3dIntransProbs}, where we see that the uniform model is one order of magnitude more likely to be permanent compared to the tree hierarchy model. Intransitive systems contain a fixed point in the interior of the simplex which is either stable or unstable, as the probability of a cycle fixed point is a zero-measure set. Also, the conditions that determine stability of the fixed point are symmetric, which shows that the probability of permanence should be $1/2$ of the probability of intransitivity. The reported numerical results are in agreement with this theory.  


\begin{table}[!htb]
  \centering
  \begin{tabular}{cc|c|c|}\hline
    \multicolumn{2}{|c|}{\multirow{2}{*}{Probability}}
                      & Pr(Intransitivity) & Pr(Intrans. \& Perm. interior FP) \\
    \multicolumn{2}{|c|}{}
                      & (Table~\ref{tabIntransConds}, row 1) & (Table~\ref{tabIntransConds}, row 2)\\ \hline
    \multicolumn{1}{ |c  }{\multirow{2}{*}{$\mathcal{E}_i$ model} } &
    \multicolumn{1}{ |l| }{Uni$(0,1)$, Indep.} & 23$\times10^{-4}$ & 11$\times10^{-4}$  \\ \cline{2-4}
    \multicolumn{1}{ |c  }{}                        &
    \multicolumn{1}{ |l| }{Uni$(0,1)$, Hier.} & 4.3$\times10^{-4}$ & 2.1$\times10^{-4}$\\ \cline{1-4}
  \end{tabular}
  \caption{ Probability of permanence in intransitive scenario by the criteria in Table~\ref{tabIntransConds}. }\label{tab3dIntransProbs}
\end{table}


\subsection{Comparison of coexistent systems}\label{SecResComp}
We have so far investigated how likely the system is to exhibit certain properties on the level of pairs of species, given certain assumptions on the interaction terms.  
We now return to the original question, which was: {If we have qualitative information about the pair-wise dynamics of three species, what can be said about the likelihood of coexistence in the three species community?}


This question can be answered by looking at 
the probability of permanence conditioned on existence of the system. For example, if we know that among the three species two of them coexist in pairs, how likely is the system to exhibit coexistence of all three species? We investigated this for pair-wise coexistence and intransitive triplets and the results are collected in Table~\ref{tab3dCondProb}. First, we note that there is a pattern to the pair-wise coexistent systems in that the conditional probability of permanence is increasing with the number of coexistent pairs. This pattern is the same for both the independent and hierarchical model for the interactions parameters. We also note that the conditional probability of permanence in triplets with three coexistent pairs is by far larger than the probability for systems with two or one coexistent pairs. Finally, although intransitive systems are unlikely to exist when the interactions parameters are random (as shown in Table~\ref{tab3dIntransProbs}), close to half of the intransitive systems are permanent. This means that the intransitive property, and not the permanence, is the limiting factor. The converse is true for systems with one or two coexistent pairs on the boundary, as these systems are fairly likely to exist (Table~\ref{tab3d123Friends}) but have a low conditional probability of permanence.

In order to test the robustness of these results, we have calculated the corresponding probabilities with exponentially distributed parameters. This scenario describes a situation where any (finite) energy uptake is possible, but where the probability decreases exponentially with the amount of energy extracted. The rate parameter was set to 2 so as to ensure the same mean value of the random parameters as in the case with Uni(0,1)-distributed parameters. In Table~\ref{tab3dCondProb}, we see that the trend in the pair-wise coexistent systems is similar: the conditional probability of permanence is increasing with the number of fixed points on the boundary. Also, nearly half of the intransitive systems are permanent, and we note that the probability is highest for the tree hierarchy model.

\begin{table}[!htb]
  \centerline{
  \begin{tabular}{|l|ccc|c|}\hline
    \textbf{Conditional } & \multicolumn{3}{ c| }{Pair-wise coexistent and permanent} & Intransitive and \\
    \textbf{probability of} & Three pairs & Two pairs & One pair & permanent \\
    \textbf{permanence} & \multicolumn{3}{ c| }{(Table~\ref{tab3d123Friends})} & (Table~\ref{tab3dIntransProbs}) \\ \hline
    Uni$(0,\ 1)$, Indep.  & 0.549 & 8.95$\times10^{-2}$ & 17.6$\times10^{-3}$ & 0.499 \\ \hline
    Uni$(0,\ 1)$, Hier    & 0.492 & 8.30$\times10^{-2}$ & 5.22$\times10^{-3}$ & 0.498 \\ \hline\hline
    Exp$(2)$, Indep.      & 0.476 & 14.2$\times10^{-2}$ & 31.7$\times10^{-3}$ & 0.497\\ \hline
    Exp$(2)$, Hier        & 0.429 & 7.07$\times10^{-2}$ & 2.24$\times10^{-3}$ & 0.558 \\ \hline
  \end{tabular}   }
  \caption{ Probability of permanence conditional on existence of the different types of systems, collected from Table~\ref{tab3d123Friends} and \ref{tab3dIntransProbs}. In half of the cases where we have a species triplet with pairwise coexistence for all three pairs, or if the three species are intransitive, we will also have permanence for the triple ecology. It is also interesting to compare the column for 'One pair' with the last column in Table~\ref{tab3dGenPerm} where we note that even though we condition the system to have one coexisting pair, that does not increase the probability for a triple coexistence compared to no structural conditions at all. The reason for this is that we then rule out the possibility to have an intransitive triple, which we see from the table gives about a 50\% chance of coexistence. \label{tab3dCondProb}}
\end{table}

%% file: discussion.tex
\section{Discussion}\label{secDisc}



We have investigated the statistical properties of the replicator system \eqref{repSystEq} via the parameters $\mathcal{E}_{i}$, $\mathcal{E}_{ij}$ that represent the energy uptake of a species \citep{Lundh2013}. These parameters are modeled as random variables and we use two example distributions for the parameters. The Uni(0,1) distribution is chosen to investigate the properties of the system under the assumption that any energy uptake is equally likely when normalised onto the [0,1]-interval, and the Exp(2) distribution is chosen to test the robustness of the results. Other possible distributions are log-normal and the gamma distributions, where a certain interval for the extracted energy may be specified. 


Two schemes are used for the random interactions parameters the independent where any species is allowed to extract any amount of energy from any metabolite and, in particular, may extract more energy from a derived metabolite than from the primary nutrient. This is the most general case of cross-feeding, where different species may be specialised on different nutrients. To investigate the case where a species is not able to extract more energy further down in the metabolic chain, the tree hierarchy model which satisfies the condition $\mathcal{E}_{ij}<\mathcal{E}_{i}$ is used as an alternative. 
In general the results are similar across both the different distributions and the two models, which suggests a certain robustness. However one trend seems to be that for both distributions the tree hierarchy model exhibits a lower probability of permanence. The reasons for this is that in the tree hierarchy model the first-order interactions ($\mathcal{E}_{i}$) are larger than the second-order interactions ($\mathcal{E}_{ij}$), which tends to suppress coexistence. This can be understood intuitively by considering the limiting case where the second-order terms tend to zero. In that case coexistence is improbable, and the species with the largest first-order term dominates no matter how small the relative advantage compared to the other species.

We have shown that triplet coexistence is more or less likely to occur in the three-species systems, depending on the pair-wise interactions of the systems. We find that nearly half of the systems with three coexistent pairs and intransitive systems are permanent, where the corresponding numbers for systems with one and two coexistent pairs are closer to one percent and ten percent, respectively. This is an important point when considering the possibilities for engineering permanent ecosystems: if an intransitive or pair-wise coexistent triplet is found, then it is likely to also be permanent. 



The connection to normal form games suggests that a similar statistical analysis could be readily applied to standard three strategy games. In that case one would have to decide upon a reasonable stochastic model for the elements of the payoff matrix. Here a normal distribution could be used since we no longer have any restriction on the signs of the payoffs. A natural extension would be to look at larger number of species (more than 3 strategies in a game theory context) and also include more levels of interaction in the fitness function (more than 2 players in the game). Such an analysis has been carried out by \citet{Gokhale2010}, and they could show that the maximum number of internal equilibria grows as $(d-1)^{n-1}$, where $d$ is the number of players and $n$ is the number of strategies. However, numerical results showed that the fraction of games (generated at random) that allow for coexistence of all strategies/species rapidly approaches zero as $d$ and $n$ grow. Here our methodology could be applied to derive criteria that could aid in generating and finding games that exhibit coexistence and permanence.  

One can compare  the results in Table~\ref{tab3dGenPerm} with \citep[Theorems 1 and 2]{Han2012}  where one can find that their random game with two players and three strategies has a probability of $\frac14$ for the existence of  an internal fixed point and a probability $\frac18$ for a {\em stable} internal fixed point. So, at least for these cases, we see that the cross-feeding set up makes it less probable for both existence and stability of fixed points than for the random matrix set up in \citep{Han2012}. The explanation for this can be found in the generation of the cross-feeding random matrix, see Equation \eqref{3dLGPayoffNormal} which consists of elements, $\lambda_{ij}$, that are not independent since $\lambda_{ij}$ and $\lambda_{ik}$ share one stochastic variable in one of the two terms of  $\lambda_{ij}=\mathcal{E}_{i}+\eta\mathcal{E}_{ji}$. This positive correlation makes it harder to find  a system where the second order metabolites are compensating enough with the first order pay-off. 

In a recent study by \citet{Huang2012} the emergence of polymorphism was studied in an evolutionary variant of the replicator equation. With a small mutation rate the payoff matrix was expanded with new random entries corresponding to the introduction of a new strategy/species into the ecosystem. When analysing coexisting triplets that emerged in this evolutionary process they found that almost all of them exhibited pair-wise coexistence, while only small fraction shwed Prisoner's dilemma like dominance when considered as pairs. That result is in line with what one would expect from our analysis. Pair-wise coexistence might be an unlikely scenario, but when it occurs then it is highly likely (compared to the other cases) to yield permanence of the triplet.

The replicator system of equations can be shown to be equivalent to the Lotka--Volterra system of ODEs \citep{Bomze1983}. Studies of a such a Lotka--Volterra system with normally distributed random interaction parameters showed \citep{Coyte2015} that cooperative interactions in a consortium are not likely to form stable permanence for a large number of species. In the model, the dependency between cooperating species that allowed cross-feeding or otherwise mutualistic groups to thrive also caused instability - when one of the cooperating species declined in numbers, others were likely to follow. 

We hope that the results derived in this paper, and future extensions of it, will be useful for experimentalists trying assemble stable microbial communities. Our results could aid in the construction of artificial communities that perform well-defined industrial functions. With the current advances in synthetic biology our predictions could also be tested by using genetic cross-feeding switches that are tuneable to achieve interaction strengths within a given range \citep{Kerner2012}. By generating a large number of different mutants one could investigate if indeed three pair-wise coexistent species are most likely to show triplet coexistence, and if the fraction of such system is approximately 50\%.

In conclusion, we think that this and similar bottom-up studies \citep{Gedeon2015}, that strive for an understanding of basic mechanisms in cross-feeding, have a lot to offer when it comes to understanding complex microbial ecosystems.

\section{Acknowledgements}
The authors would like to thank the both anonymous reviewers for their careful reading, hard questions and valuable suggestions that significantly improved this paper. 
PG would like to acknowledge funding from the Swedish Foundation for Strategic Research (Grant no.\ AM13-0046) and Vetenskapsr\aa det (Grant no.\ 2014-6095)
